\documentclass[preprint,showpacs,preprintnumbers,amsmath,amssymb,nofootinbib]{revtex4}

\usepackage{graphicx}% Include figure files
\usepackage{color}
\usepackage{dcolumn}% Align table columns on decimal point
\usepackage{bm}% bold math

\begin{document}

\preprint{APS/123-QED}

\title{Probing the creatable character of perturbed Friedmann-Robertson-Walker universes}

\author{Ramon Lapiedra}
 \email{ramon.lapiedra@uv.es}
\author{Diego S\'aez}%
 \email{diego.saez@uv.es}
\affiliation{Departament d'Astronomia i Astrof\'{\i}sica,
\\Universitat de Val\`encia, 46100 Burjassot, Val\`encia, Spain.}

\date{\today}% It is always \today, today,

\begin{abstract}
We discuss whether some perturbed Friedmann-Robertson-Walker (FRW)
universes could be creatable, i. e., could have vanishing
energy, linear momentum and angular momentum, as it could be
expectable if the Universe arose as a quantum fluctuation. 
On account of previous results, the
background is assumed to be either closed (with very small
curvature) or flat. In the first case, fully arbitrary linear
perturbations are considered; whereas in the flat case, it is
assumed the existence of: (i) inflationary scalar perturbations,
that is to say, Gaussian adiabatic scalar perturbations having an
spectrum close to the Harrison-Zel'dovich one, and (ii) arbitrary
tensor perturbations. We conclude that, 
any closed perturbed universe is creatable, and also
that, irrespective of the spectrum and properties of the
inflationary gravitational waves, perturbed flat FRW universes
with standard inflation are not creatable. Some considerations on
pre-inflationary scalar perturbations
are also presented. The creatable character of perturbed FRW universes
is studied, for the first time, in this paper.
\end{abstract}


\maketitle

\section{Introduction}
\label{intro}

In \cite{Ferrando} (see also the addenda in \cite{Ferrando-bis}),
for a wide set of non asymptotic Minkowskian space-times,
it was uniquely determined when one of these space-times can be
said to have vanishing energy, vanishing linear 3-momentum and
vanishing angular 4-momentum, that is, vanishing energy and
momenta. These vanishing values could be expected in the case of
a universe which rose from a quantum vacuum fluctuation
\cite{Tryon}\cite{Albrow}. In \cite{Ferrando}, universes of this
kind, with vanishing energy and momenta, were called `creatable
universes'.

It is very well known that, whatever the energy-momentum complex
may be, the definition of energy and momenta of the universe is
strongly dependent on the coordinate system. One must then stress
the uniqueness reached \cite{Ferrando} in the characterization of
the family of creatable universes, within the above wide set.
This uniqueness has been reached by using some physical criteria
to select the appropriate coordinate systems. See \cite{Garecki}
as an example of a more mathematical criterion, which leads the
author to use conformally flat coordinates in the space-time, when
it is possible.

In order to look for convenient physical criteria to select the
appropriate coordinates, one should assume that the proper energy
and momenta of any space-time representing the universe is
conserved in time. Taking into account that the Universe is
supposed to embrace everything, fluxes of energy and momenta going
out of such an entity are not possible and, consequently, the
above assumption about time conservation seems actually
appropriate. Such an assumption only has a clear physical meaning
if it is referred to a physical (proper) and universal
(synchronous) time. In other words, the conserved energy and
momenta of a universe should be calculated by using Gauss
coordinates, which involve the required time coordinate. Energy
and momenta would be then conserved in terms of Gauss time
coordinates. Because of this conservation, no global
Gauss coordinates are necessary in practise to calculate energy
and momenta. Actually, Gauss coordinates defined in the elementary
vicinity of a generic space-like 3-surface suffice.

Given a space-time, there are plenty of different Gauss
coordinate systems. Thus, we will need to make sure that 
the characterization of any perturbed FRW universe as creatable or
non creatable is independent on the
Gauss coordinate used. We will consider various of these perturbed 
universes and, then, in order to see if they are creatable,
we will apply different strategies, inspired by \cite{Ferrando}, which are
adapted to each particular case. Nevertheless, any of these
strategies will obey the following protocol:
(i) take any space-like 3-surface, $\Sigma_3$, and build the
corresponding Gauss coordinates in its elementary vicinity, (ii)
look for new Gauss coordinates
leading to an `instantaneous' 3-space metric,
$dl_0^{2}\equiv {dl_0^2(t=t_0)}$, which explicitly exhibits its
conformal character on the
boundary, $\Sigma_2$, of $\Sigma_3$ (these coordinates always exist; see
\cite{Ferrando}) and, (iii) calculate the energy and momenta of
the universe in the resulting coordinate system. Despite the still
remaining freedom in the election of these coordinates, it can be
seen that the calculated values of energy and momenta are unique
for the different cases we consider in the present paper. 
In all these cases, 
if these energy and momenta vanish, the universe is creatable,
since it can be directly seen that they vanish irrespective of both the
selected space-like 3-surface, $\Sigma_3$, and the chosen time, $t_{0}$
(energy and momenta conservation). 
This will be verified at the end of each particular case.

It has sometimes been argued (see, e. g., \cite{Garecki})
that, in a space-time which is not
asymptotically flat, the global energy and momenta
would have no physical meaning, since these quantities could never
be measured; however, it is also claimed that the energy and momenta
of any part of this space-time have physical meaning. This double
claim is not fully consistent; in fact, if energy and momenta have
physical meaning for any part of the space-time,
the corresponding global quantities should be interpreted as the limit of
the physically significant energy and
momenta of the parts, as they grow to fill the entire
space-time.

By using the Weinberg energy-momentum complex \cite{Weinberg} and
the above protocol, it was proved that the closed and flat
Friedmann-Robertson-Walker (FRW) universes ($K=0,+1$) are
creatable, whereas the open version ($K=-1$) is not
\cite{Ferrando}. This is in agreement with most of the papers on
the subject, but not with all of them (see again \cite{Garecki}).
Similar conclusions were obtained by using very different 
methods. For example, in papers \cite{Atkatz} and \cite{Vilenkin}, 
it was proved that the tunneling amplitude for creation from 'nothing'
is finite in the case of closed cosmological backgrounds and 
for a flat De Sitter unperturbed universe. These two papers concluded that the tunneling 
amplitude vanishes in the open case. 
These conclusions were obtained in the framework of particular
FRW models for tunneling.
Realistic universes with 
perturbations were not considered at all. 
The main goal of the present paper is the study of perturbed 
FRW universes without any calculation of transition amplitudes.
Instead of these calculations,
we use the above protocol to look for creatable universes 
(those with vanishing energy and momenta) without modelling any
quantum tunneling.

The observations seem to indicate that we live in a perturbed FRW
universe and, consequently, if the above protocol makes sense, at
least one of the perturbed FRW universes should be creatable. Our
results confirm this expectation, since we have proved that any
perturbed closed universe is creatable. The perturbed $K=-1$
universe does not need to be considered here because the
corresponding background is not creatable. The perturbed flat case
is studied in detail along the paper. It is not creatable, under
very general conditions strongly supported by current
observations.

This paper is organized as follows: In Sec.~\ref{sec-2}, we
consider absolutely arbitrary linear perturbations in the case
$K=+1$. The case $K=0$ is studied in Sec.~\ref{sec-3}. The total
energy due to the scalar perturbations arising in standard
inflationary models is calculated in Sec.~\ref{sec-3A} (it is
infinite). Sec.~\ref{sec-3B} contains the calculation of the same
quantity in the case of fully arbitrary gravitational waves (it
vanishes). Finally, in Sec.~\ref{sec-4}, we summarize our main
conclusions and present a general discussion, including 
comparisons between our results and methods and those of previous papers
dealing with quantum creation from 'nothing' (see references
\cite{Atkatz}--\cite{Hartle}).

Some of these results have been briefly presented, with no
calculations, in the meeting ERE-2007 \cite{Lapiedra}

Let us finish this section with some words about notation. Units
are chosen in such a way that the speed of light is $c=1$. The
gravitational constant, the scale factor, and the index of the
3-space curvature are denoted $G$, $a$, and $K$, respectively.
Symbols $t$ and $\tau$ stand for the Gauss and the conformal times
($a d\tau= dt$) and, finally, the unit vector $\hat{\mathbf{k}}$
and the modulus $k$ define a generic vector
$\mathbf{k}=k\hat{\mathbf{k}}$ in momentum space.

\section{The case of a perturbed closed FRW universe}
\label{sec-2}

The line element of the closed FRW space-time can always be
written in the form:
\begin{equation}
ds^2 = -dt^2+dl^2, \, \,dl^2 =
\frac{a^2(t)}{\left(1+\frac{K}{4}r^2\right)^2} \delta_{ij} dx^i
dx^j \, , \quad r^2 \equiv \delta_{ij} x^i x^j \, \label{FRW
metric}
\end{equation}
with $K=+1$. Here, $x^{i}$ and $t$ are global Gauss coordinates.
Moreover, the
3-space metric exhibits a conformally flat form everywhere.

Now, let us consider the case of a perturbed closed FRW universe
with scalar and tensor
perturbations \cite{Bardeen}. Then,
in the synchronous gauge
(see the Appendix),
conditions $g_{00}=-1$, $g_{i0}=0$ are satisfied. Hence, in this gauge,
Gauss coordinates are used. In terms of them,
the 3-space metric,
$dl^2$, reads as follows:
\begin{equation}
dl^2 = \frac{a^2(t)}{\left(1+\frac{1}{4}r^2\right)^2}
(\delta_{ij}+h_{ij}) dx^i dx^j, \label{perturbed FRW metric}
\end{equation}
where the $h_{ij}(t,x^{i})$ functions are such that
$h_{ij}<<1$.

Let it be a particular space-like 3-surface, $t=t_{0}$, and
particularize the above 3-space metric on this 3-surface, that is,
consider $dl(t=t_{0})^2 \equiv{dl_0^2}$. This instantaneous
3-space metric is a conformally flat 3-metric on the boundary
2-surface, $\Sigma_2$, of $t=t_{0}$ \cite{Ferrando}. Then, among
the different Gauss coordinate systems, one always can
select some ones such that $dl_0^2$ on the boundary $\Sigma_2$,
say $dl_0^2|_{\Sigma_{2}}$, shows explicitly its conformally flat
character. According to the protocol displayed in
Sec.~\ref{intro}, which can be used here to calculate the energy
and momenta of perturbed closed universes, 
we can pick up any one of these last
coordinate systems to compute the corresponding values of the
energy and momenta of our perturbed universe. We will subsequently
see that these values are unique.

Then, according to \cite{Weinberg}, the
energy, $P^{0}$, the 3-momentum, $P^{i}$, and the 4-angular
momentum, $(J^{jk},J^{0i})$, of the universe are:
\begin{eqnarray}
P^0 & = & \frac{1}{16 \pi G} \int(\partial_j g_{ij} - \partial_i
g)
d \Sigma_{2i},  \label{energy}\\[3mm]
P^i & = & \frac{1}{16 \pi G} \int(\dot{g} \delta_{ij} -
\dot{g}_{ij}) d \Sigma_{2j},
\label{three-momentum}\\[3mm]
J^{jk} & = & \frac{1}{16 \pi G} \int(x_k \dot{g}_{ij} - x_j
\dot{g}_{ki}) d \Sigma_{2i},\label{angular three-momentum}
\\[3mm]
J^{0i} & = & P^i t - \frac{1}{16 \pi G} \int[(\partial_k g_{kj} -
\partial_j g)x_i + g \delta_{ij} - g_{ij}] d \Sigma_{2j},
\label{angular time momentum}
\end{eqnarray}
where $\dot{g_{ij}} \equiv
\partial_t {g_{ij}}$, $g\equiv \delta_{ij}g_{ij}$, and where
$d\Sigma_{2i}$ stands for the integration element on
$\Sigma_2$. In the present case, the 2-surface $\Sigma_2$ is
$r=\infty$.

Now, it is obvious that the energy and
momenta of our perturbed closed FRW universe must vanish; in fact,
according to Eq. (\ref{perturbed FRW metric}), all the integrands in
the above 2-surface integrals go at least like $1/r^4$ as $r$
tends to $\infty$. (Notice that $h_{ij}$ cannot grow
indefinitely with $r$, otherwise we would not have
$h_{ij}<<1$ everywhere). Of course, this asymptotic
behavior, in terms of $1/r$, is valid for all Gauss coordinates
which preserve the perturbed character of the metric Eq.
(\ref{perturbed FRW metric}), and so, whatever be the time
parameter $t_0$, it is valid for the particular
Gauss coordinate system where the 3-metric, $dl_0^2$, shows explicitly its
conformal flat character on ${\Sigma_{2}}$.
Evidently, such a behavior of the integrands implies straightforward
that all the integrals in
Eqs.~(\ref{energy})--(\ref{angular time momentum})
vanish irrespective of the above Gauss coordinate used.
Consequently, we can state that,
whatever the properties of the
perturbations may be, perturbed closed FRW universes are
creatable.

\section{The case of a perturbed flat FRW universe}
\label{sec-3}

In the case of a flat background, we cannot invoke
the strong decaying of the 3-space
metric, as $r$ tends to $\infty$, to conclude that energy and momenta
vanish; namely, arguments similar to those of the last section
do not apply.
Furthermore, one could erroneously think that, in the
flat case, the vanishing of the energy would come straightforward
from the cosmological principle; namely, from the assumed
statistical homogeneity
when averaging on large enough volumes. Actually, as we
will see in detail along this section, the values of the energy
and momenta of the universe
depend on both the statistical character and the spectra
of the perturbations.

\subsection{Scalar perturbations}
\label{sec-3A}

Let us first consider the case of scalar perturbations
\cite{Bardeen} in a flat FRW universe. In any synchronous gauge
(Gauss coordinates), the perturbed 3-space metric is
$g_{ij}=a^{2}(\tau)(\delta_{ij}+h_{ij})$ and, according to
\cite{Ma}, the metric perturbation, $h_{ij}$, can be expanded in
scalar harmonics (plane waves) as follows:
\begin{equation}
h_{ij}(\mathbf{x},\tau)=\int d^{3}k e^{i\mathbf{k}.\mathbf{x}} h_{ij}(\mathbf{k},\tau)
= \int d^{3}k
e^{i\mathbf{k}.\mathbf{x}}[{\hat{\mathbf{k}}}^{i}{\hat{\mathbf{k}}}^{j}h(\mathbf{k},\tau)
+({\hat{\mathbf{k}}}^{i}{\hat{\mathbf{k}}}^{j}-\frac{1}{3}\delta_{ij})6\eta(\mathbf{k},\tau)] \ ,
\label{original metric}
\end{equation}
where two functions, $h(\mathbf{k},\tau)$ and
$\eta(\mathbf{k},\tau)$, defined in momentum space, have been
introduced. Notice that, in order to have a real value for
$h_{ij}$, functions $h$ and $\eta$ must satisfy the conditions
$h(-\mathbf{k})=-h^{\star}(\mathbf{k})$ and
$\eta(-\mathbf{k})=-\eta^{\star}(\mathbf{k})$.

According to the above protocol to calculate energy
and momenta, we must
use a new Gauss coordinate system, in which, the transformed
components of the instantaneous 3-space metric,
$g'_{ij}(t=t_0)$, show explicitly its conformal
flat character
on the boundary $r=\infty$. In the present case, we can do
more than this since our instantaneous 3-space metric is a
conformally flat metric everywhere on the space-like 3-surface
$t=t_{0}$ (not only on the 2-surface $r=\infty$).
In order to prove this statement, let us work in the
$\mathbf{k}$-space. Given a $\mathbf{k}$-value, let us consider the
infinitesimal coordinate transformation
\begin{equation}
x^{i}={x}'^i+{e}^{i}(\mathbf{x},\mathbf {k}), \label{new
coordinates}
\end{equation}
which leads to the following equation:
\begin{equation}
{h}'_{ij}(\mathbf{k})e^{i\mathbf{k}.{x}}=h_{ij}(\mathbf{k})e^{i\mathbf{k}.{x}}+
\partial_{i}e_{j}+\partial_{j}e_{i} \ .
\label{new metric}
\end{equation}
If we then choose
$e^{i}(\mathbf{x},\mathbf{k})=f(\mathbf{k})e^{i\mathbf{k}.\mathbf{x}}\hat{\mathbf{k}}^{i}$,
where $f(\mathbf{k})$ is defined by the relation
\begin{equation}
(h+6\eta+2a^{2}k^{2}f)_{t=t_{0}}=0 \ ,
\end{equation}
equation (\ref{new metric}) leads to the following
new Fourier component ${h}'_{ij}(\mathbf{k})$ for $t=t_{0}$
\begin{equation}
{h}'_{ij}(\mathbf{k})_{t=t_{0}}\equiv{{h}'_{ij}(\mathbf{k})}_{0}=
-2\eta_{0}(\mathbf{k})\delta_{ij}, \label{new-h}
\end{equation}
where $\eta_0(\mathbf{k})\equiv{\eta(\mathbf{k},t=t_0)}$. Equation (\ref{new-h})
implies that the instantaneous metric $g_{ij}(t=t_{0})$ is
conformally flat everywhere on the 3-space $t=t_{0}$, as we
wanted to prove.

To calculate the energy, we must insert the new components of the
3-space metric ${g}'_{ij}(t=t_{0})=a^2_0[\delta_{ij}+{h}'_{ij}(t=t_{0})]$
in the integral of Eq.~(\ref{energy}), which must be performed on the
2-surface $r=\infty$. This integral is to be calculated in the new
coordinates ${x}'^{i}$; however, in practice, the old coordinates, $x^{i}$,
can be used in first order calculations; in fact, since the
integrand in Eq.~(\ref{energy}) trivially vanishes for the background metric
(a flat FRW universe), this integrand is a first
order quantity in the perturbed universe and, consequently, the
integral can be evaluated
irrespective of the coordinate system
(the differences between the integrands in the old and new coordinates
are second order
quantities to be neglected in any linear approach).
In short, to first
order, metric perturbations and its derivatives can be
calculated in terms of the old coordinates, and
the old 2-surface element, $d\Sigma_{2i}$, can be used instead of
the new one.
Thus, according to Eq. (\ref{new-h}), the total energy
$P^{0}$ can be written as follows:
\begin{equation}
P^{0}=\lim_{r\rightarrow \infty}\frac{a_0^2 r^{2}}{8\pi G}\int
d\Omega n_{i}\partial{_i}\int dk^{3}\eta_{0}
e^{i\mathbf{k}.\mathbf{x}}, \label{special energy}
\end{equation}
where $d\Omega$ is the solid angle element in spherical
coordinates, i. e., $d\Omega=\sin\theta d\theta d\phi$, and where
$n_{i}\equiv{x^i/r}$.

It is noticeable that, according to the above
expression of $P^0$, the energy of the universe does not depend on
the function $h$ in Eq.~(\ref{original metric}). It only depends on
the function $\eta$, that is, on the traceless part of the
perturbed metric (\ref{original metric}).

We assume statistical isotropy, which characterizes
cosmological processes as, e. g., standard inflation.
Since there are no privileged directions,
the power spectrum of
$\eta(\mathbf{k})$, namely, the function
$P_{\eta}(\mathbf{k})=\langle |\eta(\mathbf{k})|^{2} \rangle $ is
independent on $\hat{\mathbf{k}}$. It only
depends on $k$. Three $k$-intervals can be distinguished (see the
Appendix). The interval ($0,k_{min}$) involving pre-inflationary
perturbations
with some unknown power spectrum $P_{\eta 1}(\mathbf{k})$,
the ($k_{min},k_{max}$) interval with inflationary
perturbations evolving outside the horizon, whose
spectrum is $P_{\eta 2}(\mathbf{k})$, and the
the interval ($k_{max},Â\infty$) with inflationary perturbations
which have reentered the horizon; in this case,
the spectrum is denoted $P_{\eta 3}(\mathbf{k})$.

Inflationary perturbations are all Gaussian
in the evolution period under consideration (before entering in nonlinear processes,
see the Appendix);
hence, if pre-inflationary perturbations are also assumed to be Gaussian
(the non Gaussian case is discussed below),
functions $\eta $ and $h$, as well as the Fourier
transform of any physical quantity
(density contrasts, pressure, and so on) are complex
functions with random phases \cite{Peebles}
for $0 < k < \infty$ and, consequently,
the $\eta $ values necessary to calculate $P^0$
--from Eq.~(\ref{energy})--
can be written in the form
\begin{equation}
\eta(\mathbf{k}) = \frac {1}{\sqrt{2}} \Big[P_{_{\eta}}(k)\Big]^{1/2}
\Big[y_1(\mathbf{k})+iy_2(\mathbf{k})\Big] \ , \label{gaussian eta}
\end{equation}
where
$y_1(\mathbf{k})+iy_2(\mathbf{k})$ are the phases, which must be assigned
taking into account
the relation $\eta(-\mathbf{k})=-\eta^*(\mathbf{k})$ and the
relations:
\begin{equation}
\langle y_1(\mathbf{k}) \rangle =\langle y_2(\mathbf{k}) \rangle =0 \ ,
\label{c1}
\end{equation}
\begin{equation}
\langle y_1(\mathbf{k}) \, y_2({\mathbf{k}}')\rangle =0 \ ,
\label{c2}
\end{equation}
\begin{equation}
\langle y_n(\mathbf{k}) \, y_n({\mathbf{k}}')\rangle
=\delta^3(\mathbf{k}-\mathbf{\mathbf{k}}') \ ,
\label{c3}
\end{equation}
where index $n$ run from $1$ to $2$, vectors
$\mathbf{k}$ and ${\mathbf{k}}'$ are arbitrary momenta, and
the averages are to be performed in the set of the universe realizations.
In the last equation, $\delta^3$ is the three
dimensional Dirac $\delta$-distribution.
Equations~(\ref{c1})--(\ref{c3}) ensure (central limit theorem)
that the distribution
of $P^0$ values corresponding to all the possible
realizations of the universe is Gaussian.

By using the relation
\begin{equation}
\int d\Omega n_i e^{i\mathbf{k}.\mathbf{x}}=\frac{4\pi
i}{kr}\Big(\frac{\sin kr}{kr}-\cos kr\Big)\mathbf{\hat{k}}_i \ ,
\label{integral}
\end{equation}
which can be easily obtained, plus Eqs. (\ref{special energy}) and
(\ref{gaussian eta}), one easily finds:
\begin{equation}
P^0=-(a_0^2/2G)\lim_{r\rightarrow \infty}r \sum_{j=1}^{3} \int_{j} d^3
k  \Big[P_{\eta j}(k)\Big]^{1/2}
\Big(\frac{\sin kr}{kr}-\cos kr\Big) y_1(\mathbf{k}) \ ,
\label{xxx}
\end{equation}
where the index $j$ number the three intervals defined above. The
volume integrals in momentum space are extended to the regions
$k<k_{min}$, $k_{min}<k<k_{max}$, and $k>k_{max}$, in cases j=1,
j=2, and j=3, respectively. In order to derive the last equation,
we have used the relation $\eta(-\mathbf{k})=-\eta^*(\mathbf{k})$
and, consequently, the energy $P^0$ appears to be real valued as
it must be. It is also important that, as a result of the same
relation, quantity $P^0$ only depends on the random variable
$y_1(\mathbf{k})$; whereas $y_2(\mathbf{k})$ becomes fully
irrelevant in our calculations.

Taking into account Eq.~(\ref{xxx}) and the fact that
$\langle y_1(\mathbf{k}) \rangle$ vanishes, one easily finds the relation
$\langle P^0 \rangle=0$. Of course, the average is performed for
all the possible realizations of the universe, which
correspond to distinct $y_{1}(\mathbf{k})$ values but to the same power
spectrum. Since the phases have been chosen in such a way
that the $P^0$ values are normally distributed, we must now
calculate the variance, $\langle (P^{0})^{2} \rangle$, which fully
characterize this normal (Gaussian)
distribution with vanishing mean. From Eq.~(\ref{xxx}), one easily finds a formula
for $(P^{0})^{2}$ which is the addition of six different terms. Three of
them involve products of integrals corresponding to
distinct $j$-values (hereafter crossed terms).
Since two $\mathbf{k}$-vectors corresponding
to distinct $j$ values cannot coincide, equation~(\ref{c3}) implies
that, after averaging, the crossed terms vanish.
From the remaining three terms plus Eq.~(\ref{c3}) one easily obtains
the following variance:
\begin{equation}
\langle (P^{0})^{2} \rangle=(a_0^4/4G^2)\lim_{r\rightarrow \infty} r^2 \sum_{j=1}^{3} \int_{j}
d^3k P_{\eta j}(k) \Big(\frac{\sin kr}{kr}-\cos kr \Big)^2 \ .
\label{second special energy}
\end{equation}
Each of the three terms of this last formula is positive (it is the square
of one of the three integrals $j=1,2,3$) and,
consequently, if one of them diverges, the variance
$\langle (P^{0})^{2} \rangle$ diverges.

In the case of inflationary adiabatic
scalar perturbations which evolve outside the horizon
in the radiation dominated era ($k_{min}<k<k_{max}$),
the solution of the linearized
Einstein field equations
can be found in Ma and Bertschinger \cite{Ma} (see the Appendix
for more details). In the synchronous gauge, function $\eta $
evolves as follows:
\begin{equation}
\eta=C+\frac{3}{4}\frac{D}{k\tau}+\alpha Ck^{2}\tau^{2} \ ; \label{eta}
\end{equation}
in this equation, $\alpha$ is a pure number (see the Appendix),
whereas $C$ and $D$ are
dimensionless integration constants. That is to say, $C$ and $D$
cannot depend on time, but they may depend on $\mathbf{k}$. Notice that
here, as well as in \cite{Ma}, the components of $\mathbf{x}$ and $\mathbf{k}$
are dimensionless, since they are comoving coordinates and their
associated momenta. We see that $\eta$ is the addition of three different
modes: one of them grows like $\tau^{2}$, another one is
constant, and the third one decreases like $\tau^{-1}$.

Let us now study the behavior of the integrals involved in
Eq. (\ref{second special energy}) as $r$ tends to $\infty$.
We begin with one of the contributions to the $j=2$ integral
(inflationary Gaussian perturbations
with super-horizon sizes). It is the contribution
due to
the constant mode $C(\mathbf{k})$ in Eq.~(\ref{eta}). This
contribution is
hereafter denoted $\langle (\tilde{P}^{0})^{2} \rangle$.
The spectrum $P_{_{C}}(k)$ of the constant mode is defined by
the relation
\begin{equation}º
\Big[P_{_{C}}(k)\Big]^{1/2}=Ak^n, \label{modulus C}
\end{equation}
where $A$ is a normalization constant and the spectral index
is $n=-(3/2)-\beta$. According to Eq.~(\ref{ck2}) in the Appendix,
constant $\beta = (1-n_{s})/2$ is positive and
small as compared to unity.
From the spectrum (\ref{modulus C}), one easily finds the following
variance:
\begin{equation}
\langle (\tilde{P}^{0})^{2} \rangle=(\pi a_0^4 A^2/G^2)\lim_{r\rightarrow \infty}
r^{2+2\beta} \int_{rk_{min}}^{rk_{max}} dy\Big(\frac{\sin y}{y}-\cos
y)^2\Big)/y^{1+2\beta},
\label{yyy}
\end{equation}
where $y\equiv kr$. For any $k$, the new variable $y$ tends to infinity as
$r$ does. On account of this fact, it is easily
proved that the integral in Eq.~(\ref{yyy}) goes just like
the power $r^{-2\beta}$ as $r$ tends to infinity. Hence,
$\langle (\tilde{P}^{0})^{2} \rangle$ diverges as $r^{2}$.

Let us now consider the growing and decaying terms appearing
in Eq.~(\ref{eta}). It is obvious that the contributions
of these time dependent terms to the energy
cannot compensate the infinite value of
$\langle (\tilde{P}^{0})^{2} \rangle$ corresponding to the constant mode, which implies that
the total contribution to $\langle (P^{0})^{2} \rangle$ due to the
$j=2$ $k$-interval cannot become finite.
More precisely, from Eq.~(\ref{eta}) we find six
distinct terms
contributing to the $j=2$ integral. The term used in the previous calculation
does not depend on time, whereas the remaining ones are time dependent.
By this reason, the variance $\langle (\tilde{P}^{0})^{2} \rangle$
is proportional to $a^{4}$ for the time independent mode, whereas
it exhibits other time dependences in the remaining
cases. In all, compensation of the resulting terms (with distinct time
evolutions) to give a conserved finite total energy is not possible,
which ensures that
the entire contribution $j=2$ to $\langle P^{0})^{2} \rangle$ is infinite.
Moreover,
this positive contribution cannot be compensated by those of the $j=1$ and $j=3$ cases
to give, finally, a finite global value for $\langle P^{0})^{2} \rangle$, the reason being that,
as explained above, the $j=1$ and $j=3$ contributions are also positive.
All in all, the following relation holds:
\begin{equation}
\langle (P^{0})^{2} \rangle=\infty
\label{infif}
\end{equation}
and, consequently, we have a Gaussian statistical
distribution of $P^0$ values with zero mean and infinite variance.
Since the Gaussian probability density is
\begin{equation}
P(P^{0})= \frac {1} {\sqrt{2 \pi \langle (P^{0})^{2} \rangle}}
e^{-(P^{0}-\langle P^{0} \rangle)^{2}/2\langle (P^{0})^{2} \rangle} \ ,
\label{denpro}
\end{equation}
it is evident that, in our case,
the probability
of any particular finite value of $P^0$ vanishes and,
as a
consequence, we can say that, in the flat perturbed FRW universe
under consideration,
the contribution of the inflationary
scalar
perturbations to the total energy is
infinite. Nevertheless, in the standard inflationary
paradigm, there are also tensor perturbations, whose contribution
to the total energy of a flat perturbed universe
is calculated in next section.

Let us now discuss the case of non-Gaussian pre-inflationary perturbations.
These perturbations are assumed to be
generated in some process fully independent on inflation and, as explained in the Appendix,
they are assumed to be significant only in the interval $(0,k_{min})$; hence,
Eqs.~(\ref{c3}) are satisfied except for
pairs of vectors $\mathbf{k}$ and ${\mathbf{k}}'$ whose modulus
are both inside the interval ($0,k_{min}$). Therefore,
there are no crossed terms in the
development of $\langle (P^{0})^{2} \rangle$ (see above).
As in Eq.~(\ref{second special energy}), the variance $\langle (P^{0})^{2} \rangle$ is the
addition of three positive terms. Terms $j=1$ and $j=2$ have the same
form as in the Gaussian case, whereas the term $j=3$ would
be different. In this situation, the term $j=2$ diverges (same
discussion as above) and, consequently, Eq.~(\ref{infif}) holds
(which strongly suggests a non creatable universe).
However, the distribution of $P^{0} $ values is not Gaussian in this case,
and its probability density should be calculated for each
particular non Gaussian model. Since Eq.~(\ref{denpro}) does not apply,
the meaning of an infinite variance
is now less clear. By this reason, in order to properly prove that
flat universes with standard inflation are not creatable, we prefer a
general argument, which proves that,
if $\langle (P^{0})^{2} \rangle$ diverges in the interval
($k_{min},\infty $), namely, if the total energy of the
inflationary Gaussian scalar perturbations is infinite,
the energy of the universe cannot vanish
whatever the properties of the non-Gaussian pre-inflationary perturbations
may be; namely, the universe is not creatable.
This is trivially proved taking into account that, in any admissible universe,
the total energy
must be a conserved quantity (see Sec.~\ref{intro})
which vanishes in the creatable case.
In fact, a vanishing energy after inflation is only possible
if the pre-inflationary energy is infinite and it exactly compensates
the infinite energy associate to the scalar inflationary
perturbations. However, such a universe would have an infinite energy
before inflation and a vanishing one after this process, which
is not compatible with the required energy conservation.

Now, before ending this Section, we raise some
comments about the consistence of the main result: the
infinite value of the energy we have found.

First of all, although the infinite energy has been formally
obtained for a given value of $t_0$ and a certain $\Sigma_3$, 
it is obvious, by
following the implementation of the protocol we have used,
that this $t_0$ value and the choice of
$\Sigma_3$ are both arbitrary.

Finally, we could perform a conformal coordinate
transformation and still retain the explicit conformal flat form
of the 3-space metric on $\Sigma_3$ (remember that, according to
Eq.~(\ref{new-h}), in the present case, the 3-space metric is a
conformally flat one all over $\Sigma_3$). But, trivially, the
above infinite value of the energy does not depend on the conformal
transformations which are pertinent here: those belonging to the
translation, rotation, and dilatation subgroups of the conformal
group. For all these coordinate transformations the
energy remains infinite.

\subsection{Tensor perturbations}
\label{sec-3B}

In this section we are concerned with tensor
perturbations evolving in a flat background. In such a case,
the instantaneous 3-space line element is:
$g_{ij}=a_0^2(\delta_{ij}+h^{^{T}}_{ij})$. In Fourier space, we can write
\begin{equation}
h^{^{T}}_{ij}(\mathbf{k})=H(\mathbf{k},\tau_0)\epsilon_{ij}
(\mathbf{\hat{k}}),
\label{tensor perturbation}
\end{equation}
where the quantities $\epsilon_{ij}$ satisfy the conditions
given in the Appendix.

Let us calculate the energy, $P^0_{_{T}}$, of these tensor perturbations.
According to our protocol, calculations must be performed in a
new coordinate system, in which the instantaneous 3-dimensional
metric
explicitly exhibits its conformally flat character on the boundary 2-surface
$r=\infty$. For each value of $\mathbf{k}$, let us consider a
coordinate transformation of the form (\ref{new coordinates}).
Functions $e^{i}(\mathbf{x},\mathbf {k})$ must be chosen in such a way that,
in the new
coordinates ${x}'^i$, the line element on the surface $r=R$, where $R$ is
an arbitrary constant (at the end of our calculations, this constant will tend to $\infty$),
has the form
\begin{equation}
{h^{^{T}}}'_{ij}(\mathbf{x})\Big|_{r=R}=f(\mathbf{n},\mathbf{k})
e^{iR\mathbf{k}.\mathbf{n}}\delta_{ij} \ ,
\label{conformal metric}
\end{equation}
where ${h^{^{T}}}'_{ij}(\mathbf{x}) = {h^{^{T}}}'_{ij}(\mathbf{k})
e^{i\mathbf{k}.\mathbf{x}} $ is the metric perturbation in the
space-like hypersurface $\tau = \tau_{0} $ corresponding to the
fixed mode $\mathbf{k}$, and where function
$f(\mathbf{n},\mathbf{k})$ is the conformal factor of the metric
on $r=R$. As it has been said above, there always exists a family
of coordinate systems in which Eq.~(\ref{conformal metric}) is
satisfied \cite{Ferrando}. The energy and momenta can be
calculated in any of these coordinates. With the appropiate $e_i$
functions, from Eq.~(\ref{new coordinates}), one easily finds the
relation
\begin{equation}
{h^{^{T}}}'_{ij}(\mathbf{x})= h^{^{T}}_{ij}(\mathbf{x})
+\partial_i e_j+\partial_j e_i \ , \label{h prime}
\end{equation}
which is equivalent to Eq.~(\ref{new metric}).

The energy, $P^0_{_{T}}$, is calculated by using
Eq.~(\ref{energy}) and the metric perturbation components
${h^{^{T}}}'_{ij}(\mathbf{x'})$; nevertheless,
in the linear approach we are using
(see Sec.~\ref{sec-3A} for details),
the following approximations can be performed: (i) write these
components in terms of the old
coordinates $x^{i}$, (ii) perform the derivatives with respect to $x^{i}$, and
(iii) use the old 2-surface element,
$d\Sigma_2^{i}$, instead of $d{{\Sigma}'}_2^{i}$. Hence,
we can write
\begin{equation}
P^0_{_{T}}=(a_0^2/16\pi G)\int \partial_j\Big[{h^{^{T}}}'_{ij}(\mathbf{x})-\partial_i
{h^{^{T}}}'(\mathbf{x})\Big] d\Sigma_2^i \ .
\end{equation}
Taking into account Eq.~(\ref{h prime}),
this last equation can be rewritten as follows:
\begin{equation}
P^0_{_{T}}=(P^0_{_{T}})_H+(a_0^2/16\pi G) \int
\Big[\partial_j(\partial_i e_j+\partial_j e_i)-2\partial_i
\partial_k e_k\Big] d\Sigma_2^i \ ,
\end{equation}
where $(P^0_{_{T}})_H$ is the energy corresponding to
the first term of the r.h.s. of Eq.~(\ref{h prime}), whose
Fourier transform, $h^{^{T}}_{ij}(\mathbf{k})$, is
given by Eq.~(\ref{tensor perturbation}).
As a result of the conditions satisfied by the quantities $\epsilon_{ij}$
(see the Appendix), it is easily proved that
the term $(P^0_{_{T}})_H$ vanishes. Hence,
\begin{equation}
P^0_{_{T}}=(a_0^2/16\pi G)\int (\partial_j\partial_j
e_i-\partial_i\partial_j e_j) d\Sigma_2^i \ ,
\end{equation}
and, finally, the Gauss theorem allow us to write the surface
integral in the last equation as a vanishing volume integral 
whatever the functions  $e_i$ may be and, in particular, for the functions
leading to Eq. (\ref {conformal metric}). Hence, we have proved that
the energy associated to any distribution of gravitational waves
(propagating in a flat universe) vanishes; namely, the equation
\begin{equation}
P^0_{_{T}}=0.
\end{equation}
is satisfied.

Therefore, according to our protocol, we conclude that the energy
$P^{0}_{_{T}}$ due to tensor perturbations of a flat FRW vanishes.
Notice that this conclusion does not depend either on the spectrum
$P_{_{T}}(k)$ of the tensor perturbations (see the Appendix), or
on the statistical character of the distribution of these
perturbations.

Notice again that similarly to what has been explained to the
end of Sec.~\ref{sec-3A}, the resulting vanishing energy
does neither depend on the chosen value of the $t_0$ parameter,
nor on the choice of $\Sigma_3$.

Since the energies due to scalar and tensor perturbations add,
realistic perturbed universes including inflationary scalar modes
have in all an infinite energy and so are not creatable. The presence of arbitrary
tensor perturbation (zero energy) is irrelevant.
Coming back to the end of Sec.~\ref{sec-3A}, we see now that the energy of the
scalar perturbed closed FRW universes is infinite irrespectively of, not only 
the rotation, traslation and dilatation groups, but actually on any infinitesimal 
coordinate transformation.

\section{Conclusions and discussion}
\label{sec-4}

Our main conclusion is that perturbed flat FRW universes,
including arbitrary tensor perturbations, and the adiabatic Gaussian scalar ones
generated during standard inflation, have
an infinite energy which
is due to the scalar perturbations (see Secs. \ref{sec-3A} and \ref{sec-3B}).
Since the total energy does not vanish, perturbed flat universes are not creatable,
at least, in the framework of the standard inflationary paradigm,
which appears to be compatible with most current observations.
This conclusion implies that, among the perturbed FRW universes
undergoing ordinary inflation,
only the closed ones are creatable and, consequently, 
the slightly inhomogeneous universe where
we live should be closed.

It is generally believed that, in classical terms, there is no
any way to decide if our universe is flat or closed, at least, if
the curvature is small enough. An exception can be found in
\cite{Barrow}, where it is claimed that the {\em spiral geodesic
effect} could be used to decide, observationally, whether we live
either in a flat or a closed perturbed FRW universe. Another
different method to distinguish between flat and closed perturbed
universes arises from this paper. It is not directly based on
observations. In our case, the creatable character of closed, and
flat models compatible with observations, is studied according to
the protocol described in Sec.~\ref{intro}. 
The closed universes are creatable whatever the
linear perturbations may be. This is a robust conclusion which
privileges closed models against the flat ones. In the $K=0 $
case, we have studied the most accepted model based on standard
inflation. It does not appear to be creatable; nevertheless, this
conclusion is not valid whatever the perturbations may be. On the
contrary, it is based on some assumptions and, consequently, it
must be revised if some of such assumptions are modified in
future. Particular attention deserve our hypothesis about: (i) the
statistical isotropy of the universe and, (ii) the adiabatic and
Gaussian character of any scalar perturbation in the
post-inflationary era for $k_{min} < k < \infty$. Statistical
isotropy has been recently questioned \cite {Jaffe} \cite {Ghosh},
and isocurvarture and (or) non Gaussian perturbations (based on
cosmic strings, pre-inflationary processes and, so on) are not
completely forbidden for $k_{min} < k < \infty$. Perturbed flat
universes violating condition (i), and (or) condition (ii), and
(or) any other possible condition, require particular studies if
one wants to probe its creatable character.

The subject of the creatable character of the (non
perturbed) FRW universes has been considered in two papers in the
first eighties \cite{Atkatz}\cite{Vilenkin} (there is also related
work in \cite{Coleman} \cite{Hartle}). In
\cite{Atkatz}\cite{Vilenkin}, the authors discussed the
possibility that the Universe could have arisen by quantum
tunneling from `nothing'. In \cite {Vilenkin}, a cosmological model
is proposed in which the Universe is created by quantum tunneling
from `nothing' into a particular closed FRW model. Furthermore, in
\cite{Atkatz}, the authors find that within the context of FRW
models, only the spatially closed and the flat de Sitter universes
can originate in this manner, because they find that a finite
tunneling amplitude exists only from initial spaces with finite
three-volume (on the Euclidean section).

In the absence of perturbations, the 
results in the present paper essentially agree with those 
of \cite{Atkatz}--\cite{Hartle}, since we find with many 
other authors (including paper \cite{Ferrando})
that the closed and
flat FRW models have vanishing energy and momenta and so,
according to our terminology, this kind of universes would be
creatable. Nevertheless, the method and the scopes of our work are
very different from those of papers \cite{Atkatz}--\cite{Hartle}.

First of all, in these papers, the authors considered precise
quantum mechanisms to originate our classical universe from a
quantum one; however, we assume that the produced classical
universes must have vanishing energy and momenta and, then, we 
apply a definite
protocol to decide whether a given classical universe (in the
present case, some perturbed FRW models) has or not vanishing
energy and momenta. Since we expect that any reasonable quantum
process could not produce a universe with non
vanishing energy, we have called these universes with vanishing
energy and momenta {\em creatable} universes. In the above quoted
papers, the authors assume that energy and momenta can only be 
properly defined in
asymptotic Minkowskian universes and, consequently, they could not
follow our line of research. Nevertheless, in the present paper,
we have been able to define the energy and momenta of some non
asymptotic Minkowskian universes in a consistent and unambiguous
way (the basic idea supporting our procedures). Asymptotically, 
perturbed and non perturbed FRW universes appear to be conformally
flat, and this fact has allowed a definiton of energy and momenta which is
a generalization of that used in the case of asymptotic Minkowskian 
space-times. In both cases, we are constrained to calculate the 
energy and momenta in appropriate cordinate systems making explicit 
a certain form of the spatial metric and, then, showing how cooordinate 
transformations preserving this form do not alter the energy 
and momenta.                           

Furthermore, whereas in papers \cite{Atkatz}--\cite{Hartle} 
only exact FRW universes were studied,
we have considered realistic perturbed FRW universes
(a related study considering departures from FRW geometries 
was proposed in \cite{Atkatz}). In this way, we have found that
perturbed closed FRW universes are creatable, but perturbed flat
FRW universes, in the framework of standard inflation, are not.

Notice that, attending the different criteria put forward to
define a universe as creatable (to have a finite tunneling
amplitude, in the case of the quoted authors, or to have vanishing
energy and momenta, in our case), it is not obvious that both
definitions must lead to the same conclusions.
Comparison is possible for unperturbed FRW models (the
quoted authors have not studied the perturbed ones). There is full
agreement in the closed case, but for
non perturbed flat models, there is some discrepancy, since the
quoted authors find them creatable only in the de Sitter subcase,
whereas we find all them creatable. 
Nevertheless, this discrepancy
has perhaps a non significant meaning, since when we have
considered a realistic perturbed flat FRW model, we have found that
this model is not creatable. This result shows that the exact flat
FRW universe is an unstable one to our concerning. In the case of
perturbed FRW universes, the comparison with the results of these
authors is not possible since, as we have said, they have not
considered the case.

\begin{acknowledgments}
We would like to thank J. J. Ferrando and J. A. Morales for
valuable discussions. This work has been supported by the Spanish
Ministerio de Educaci\'on y Ciencia, MEC-FEDER project
FITS2006-06062.
\end{acknowledgments}

\appendix*
\section{Scalar and tensor inflationary perturbations in a flat background}

Our protocol begins with the use of Gauss coordinates, in which,
the metric has the form:
\begin{equation}
ds^{2}=-dt^{2}+dl^{2}, \,\,\,\,\,\, dl^{2}=g_{ij}dx^{i}dx^{j} \ ,
\end{equation}
this means that, in Gauss coordinates, the conditions
\begin{equation}
g_{00}=-1, \,\,\,\,\,\, g_{0i}=0
\label{gs}
\end{equation}
are satisfied. In the case of a FRW universe with scalar perturbations,
conditions (\ref{gs}) define the so-called synchronous gauge.
In this gauge, a detailed study about the evolution of scalar perturbations
in a flat FRW universe can be found in reference \cite{Ma}.
The authors of that paper (Ma and Bertschinger) studied scalar perturbations
in a rather general FRW flat universe containing
baryons, cold dark matter (CDM), neutrinos,
and radiation. They also studied the evolution in the
longitudinal gauge. In the synchronous gauge, Ma and Berstchinger
expanded the most general scalar metric perturbation
as it is done in Eq.~(\ref{original metric}).
This expansion
only involves two arbitrary functions: $\eta(\mathbf{k}, \tau)$ and
$h(\mathbf{k}, \tau)$.

In the longitudinal gauge one can write:
\begin{equation}
ds^{2}=a^{2}(\tau)[-(1+2\psi)d\tau^{2}+(1+2\phi)\delta_{ij}dx^{i}dx^{j}] \ ,
\label{lon}
\end{equation}
where only two arbitrary functions are necessary to describe
the most general scalar perturbation. Fourier expansions of
$\psi $ and $\phi $ involve the coefficients $\psi (\mathbf{k}, \tau)$  and
$\phi (\mathbf{k}, \tau)$.

There are other coefficients appearing in the
expansions of physical quantities involved in the energy
momentum tensor, e. g., those corresponding to the density
contrasts of the different energy components: $\delta_{\gamma}(\mathbf{k}, \tau)$
for radiation, $\delta_{c}(\mathbf{k}, \tau)$ for CDM and so on.
All these coefficients are coupled in a complicate system of
equations (see \cite{Ma}).

Fortunately, in order to evaluate the integrals giving
the total energy and momenta
of the perturbed universe, only functions
$\eta(\mathbf{k}, \tau)$ and
$h(\mathbf{k}, \tau)$ (related to the metric) could be actually necessary;
this fact facilitates our calculations.
Moreover, only functions $\eta(\mathbf{k}, \tau_0)$
and $h(\mathbf{k}, \tau_0)$ are required, $\tau_{0} $ being
an arbitrary time.
We can say that our problem is identical to that solved in
paper \cite{Ma}, where initial conditions to solve the
equations governing the evolution of the perturbations
were calculated at a fixed
time. This time was chosen to be in the time interval limited by
electron positron annihilation
and the time at which light massive neutrinos become non relativistic.
The same choice is appropriate for us.
Ma and Bertschinger solved the evolution equations in the mentioned time
interval for linear adiabatic perturbations larger than the horizon
($k \tau <<<1$). In the synchronous gauge, these authors found:
\begin{equation}
\eta=2C+\frac {5+4R_{\nu}}{6(15+4R_{\nu})}Ck^{2}\tau^{2}, \,\,\,\,\, h=Ck^{2}\tau^{2} \ ,
\label{ini}
\end{equation}
where $C=C(\mathbf{k})$, $R_{\nu}=\bar{\rho}_{\nu}/(\bar{\rho}_{\gamma}+\bar{\rho}_{\nu})$,
and $\bar{\rho}_{\nu}$ and $\bar{\rho}_{\gamma}$ are the background energy
densities of neutrinos and photons, respectively. There is also a time decaying term
whose explicit form is given in Eq.~(\ref{eta}).

The same study was also performed in the longitudinal gauge.
Indeed, Ma and Bertschinger fixed their initial conditions by assuming that, in the
longitudinal gauge, under the assumption of statistical isotropy, the power spectrum of
the $\psi $ potential of Eq.~(\ref{lon}) is
\begin{equation}
P(\psi) \propto k^{-3} \ ,
\label{spec}
\end{equation}
(see below for comments about this choice).
How can we obtain the corresponding initial conditions in the
synchronous gauge? The answer is easily obtained from Eqs. (18)
in \cite{Ma}. One of these equations reads as follows:
\begin{equation}
\psi(\mathbf{k}, \tau) = \frac{1}{2k^{2}} \Big( \ddot{h}(\mathbf{k}, \tau)+6
\ddot{\eta}(\mathbf{k}, \tau)+ \tau^{-1} [\dot{h}(\mathbf{k}, \tau)+
6\dot{\eta}(\mathbf{k}, \tau)]  \Big)  \ ,
\end{equation}
where each dot stands for a derivative with respect to the conformal time.

Taking into account Eqs. (\ref{ini})--(\ref{spec}) and this last equation,
one easily gets
\begin{equation}
C(k) \propto k^{-3/2} \ ,
\label{ck}
\end{equation}
which is valid in the synchronous gauge. That could be our basic
condition in order to compute the integrals giving the total
energy and momenta of the universe. It is also the basic
assumption leading to the initial conditions used by Ma
and Berstchinger to solve the evolution equations in
the synchronous gauge. The resulting numerical solution gave a
very good description of both the power spectrum, $P(k)$,
of the energy density perturbations and the angular power spectrum
($C_{\ell}$ coefficients) of the CMB.

Now, a question arises: why the $\psi$-spectrum defined in Eq.~(\ref{spec})
is appropriate? Let us try to answer this.

After evolution, large enough cosmological inhomogeneities
reenter the horizon in the matter dominated era and, afterward,
it is well known that the potential
$\psi $ satisfies the equation (see \cite{Ma})
\begin{equation}
\Delta \psi \propto \delta  \ .
\label{lapla}
\end{equation}
Since $\delta $ is the total energy density contrast,
this last equation indicates that function $\psi $
plays the role of the peculiar Newtonian gravitational
potential. Finally,
Eqs. (\ref{spec}) and (\ref{lapla}) lead to
\begin{equation}
P(k) = \langle |\delta_{\mathbf{k}}|^{2} \rangle \propto k \ ,
\label{hz}
\end{equation}
which means that the spectrum of the energy density perturbations
is a Harrison-Zel'dovich (HZ) one. This result justifies the
use of the $\psi $-spectrum defined
in Eq.~(\ref{spec}).
The above HZ spectrum is only valid at times close
enough to horizon crossing, but afterward, as the inhomogeneities
evolve inside the horizon, microphysics becomes
important and this spectrum evolves toward a new one of the form
$P(k)=k/T(k)$, where $T(k)$ is the so-called transfer function.

Inflationary predictions are compatible
with a HZ spectrum $P(k) \propto k$, as well as
with an spectrum of the form
$P(k) \propto k^{n_{s}}$ having its spectral index $n_{s}$ close to unity.
Accordingly, the analysis of the data obtained by the WMAP mission
during three observation years leads to the inequality
$0.942 < n_{s} < 0.974 $. Moreover, if other observational data
(galaxy surveys, other CMB observations, and so on) are taken into account,
the resulting inequality appears to be $0.932 < n_{s} < 0.962 $
(see \cite{hin07} and \cite{sper07}).
From Eq. (\ref{lapla}) one easily proves that the
condition
$C(k) \propto k^{(n_{s}-4)/2} $ leads to a final spectrum
$P(k) = k^{n_{s}} $. On account of these considerations,
our calculations of the total energy
and momenta of the universe is based on the relation:
\begin{equation}
C(k) \propto k^{(n_{s}-4)/2}, \,\,\,\, n_{s} < 1, \,\,\,\, n_{s} \simeq 1 \ ,
\label{ck2}
\end{equation}
which coincides with Eq. (\ref{ck}) for $n_{s} = 1$.

We must emphasize that Eqs.~(\ref{ini}) and (\ref{ck2})
are only valid for adiabatic scalar perturbations
evolving outside the horizon; hence, these relations only hold
for $k < k_{max} $, where $k_{max} = 2 \pi / L_{0}$ and
$L_{0}=H^{-1}(\tau_{0})$; here, $L_{0}$ is the horizon size at
the conformal time $\tau_{0}$. Moreover, if the adiabatic
perturbations are inflationary, another inequality, $k > k_{min} $,
must be also satisfied, where  $k_{min} = 2 \pi / L_{I0}$ and
$L_{I0}$ is the size, at time $\tau_{0}$, of a region comparable
to the effective horizon
at the beginning of inflation; that it to say, $L_{I0}$ is the typical
size of the huge inflationary bubbles.
For $k > k_{max} $ the perturbations evolve inside the
horizon (where microphysics is important) and, consequently,
the spectra of super-horizon perturbations must be
modified by means of transfer functions. If
these perturbations are inflationary, they are initially Gaussian
and afterward, during the radiation dominated era (in particular,
in the period considered in paper \cite{Ma} and also in Sec.~\ref{sec-3A}),
they keep Gaussian
because nonlinear processes
leading to deviations from Gaussianity had not
developed yet. Moreover, in the interval $k_{min} < k < \infty $, we assume
that the perturbations produced during inflation are
absolutely dominant against possible residual
pre-inflationary fluctuations. The main reason is that,
at the end of inflation,
inflationary supercooling had made the pre-inflationary
radiation density negligible against the total energy of the
inflationary field and, consequently, after reheating,
the mean radiation energy is fully dominated by the energy
coming from the mean inflationary field and,
evidently, the resulting
adiabatic perturbations are associated to the
fluctuations of this dominant field, with negligible
contributions from pre-inflationary supercooled sources.
Finally, for $k < k_{min} $, the perturbations are so long
that they will have a pre-inflationary origin without any
inflationary contribution. Thus, though these pre-inflationary
perturbations can be expected to be small, their contribution
to the integral in Eq.~(\ref{special energy})
could be significant as $k$ tends to zero
and $r$ tends to infinity. Thus, as a precaution,
this interval has been also considered along the paper. Of course, it has
been taken into account that these pre-inflationary
perturbations could be non Gaussian.
All these ideas are carefully taken into account
in Secs.~\ref{sec-3A} and \ref{sec-4}.

In general, inflation produces both scalar and tensor perturbations of
the background universe. Some general considerations about
tensor perturbations are now worthwhile.
The tensor metric perturbations of a flat
universe can be written in the form:
\begin{equation}
h^{^{T}}_{ij}(\mathbf{x}, \tau) = \int d^{3} k e^{i \mathbf{k} \cdot \mathbf{x}}
h^{^{T}}_{ij}(\mathbf{k}, \tau) =
\int d^{3} k \,e^{i \mathbf{k} \cdot \mathbf{x}}
H_{_{T}}(\mathbf{k}, \tau) \epsilon_{ij}(\hat{\mathbf{k}})
\ ,
\end{equation}
where functions $\epsilon_{ij} $ satisfy the
following equations:
\begin{equation}
\epsilon_{ij} = \epsilon_{ji}, \,\,\,\, \epsilon_{ii}=0, \,\,\,\, \epsilon_{ij}k_{i} =0  \ ,
\end{equation}
which ensure that quantities $h^{^{T}}_{ij}(\mathbf{x},\tau)$ are symmetric,
traceless, and divergenceless, as it must be in the case of metric
perturbations describing gravitational waves. It is
noticeable that functions $\epsilon_{ij}$ only depend on the unit vector
$\hat{\mathbf{k}}$ and, consequently, any dependence on the scale (on $k$)
of the tensor metric perturbation is involved in the coefficient
$H_{_{T}}(\mathbf{k}, \tau)$. This scale dependence is usually fixed by defining
a new power spectrum (\cite{dur01})
\begin{equation}
P_{_{T}} (k) = k^{3} \langle |H_{_{T}}(k)|^{2} \rangle \propto k^{n_{_{T}}}   \ ,
\end{equation}
where $n_{_{T}} $ is the tensor spectral index. With this spectrum,
quantities $h^{^{T}}_{ij}(\mathbf{k})$ are proportional to $k^{(n_{_{T}}-3)/2}$.
Taking into account previous formulas and assumptions, we could
calculate the total energy and momenta of the universe
for different $n_{_{T}} $ values. The explicit form of functions
$\epsilon_{ij} $ is not necessary. Nevertheless, an explicit representation
of these quantities can be easily found from the definitions
given in \cite{hu97}.
What can we say about the spectral index $n_{_{T}} $?

The spectrum generated by most inflationary potentials has
an spectral index $n_{_{T}} \simeq 1-n_{s} $  (see reference \cite{cri93}); hence,
from previous comments about $n_{s} $-values and WMAP data, it follows
that the power spectrum of these
inflationary backgrounds of gravitational waves is very flat
(small but non vanishing $n_{_{T}}$ value).
In the presence of these gravitational waves,
the spectral index $n_{s}$ deviates from unity; by this reason
and with the essential
aim of allowing the existence of an inflationary background
of gravitational waves (which is studied in Sec.~\ref{sec-3B}), the
$n_{s}$ values used in Sec.~\ref{sec-3A} have been assumed
to be slightly smaller than unity.

\end{document}